\documentclass[preprint,12pt]{elsarticle}
\usepackage[utf8]{inputenc}
\usepackage{amsmath}
\usepackage[english]{babel}
\numberwithin{equation}{section}
\usepackage{float}

\usepackage{lineno}

\usepackage[numbers]{natbib}
\title{Efficient tau-pair invariant mass reconstruction with simplified matrix element techniques}
\author[UW]{Artur Kalinowski}
\author[UW]{Wiktor Matyszkiewicz\corref{cor1}}
\cortext[cor1]{Corresponding author. Email: w.matyszkiewicz@uw.edu.pl}
\address[UW]{University of Warsaw}

\usepackage{graphicx}
\usepackage{mathtools}
\usepackage{braket}
\usepackage{mhchem}
\usepackage[shortlabels]{enumitem}
\usepackage{hyperref}

\begin{document}

\begin{abstract}
    The quality of the invariant mass reconstruction of the di-$\tau$ system is crucial for searches and analyses of di-$\tau$ resonances. Due to the presence of neutrinos in the final state, the $\tau\tau$ invariant mass cannot be calculated directly at hadron colliders, where the longitudinal momentum sum constraint cannot be applied. A number of approaches have been adopted to mitigate this issue. The most general approach uses Matrix Element (ME) integration for likelihood estimation, followed by invariant mass reconstruction as the value that maximizes the likelihood. However, this method has a significant computational cost due to the need for integration over the phase space of the decay products. We propose an algorithm that reduces the computational cost by two orders of magnitude while maintaining the invariant mass reconstruction resolution at a level comparable to that of the ME-based method. Moreover, we introduce additional features that allow the estimation of the uncertainty of the reconstructed mass and the kinematics of the initial $\tau$ leptons (e.g.\ their momenta).
\end{abstract}


\maketitle

\section{Introduction}

The di-$\tau$ final state is an important decay mode for Higgs-like particles, for which the interaction coupling is proportional to the particle mass. So far, this is the only leptonic Higgs decay channel with firm experimental confirmation~\cite{ATLAS:2022vkf, CMS:2022dwd}. The rich decay structure of the $\tau$ lepton also allows for measurements of the $\tau$ lepton polarisation and, consequently, correlations between the polarisations of the two $\tau$ leptons. This, in turn, enables studies of the CP structure of the H$\tau\tau$ interaction~\cite{Higgs_CP, Higgs_CP_Atlas}. As the search for physics beyond the Standard Model continues, the di-$\tau$ final state is also a promising channel for searches for heavy neutral Higgs bosons~\cite{ATLAS:2020zms, CMS:2022goy} or other heavy, neutral resonances~\cite{ATLAS:2015rbx, CMS:2016xbv}. The invariant mass of the di-$\tau$ system is a key handle for selecting a clean sample of signal events.

The $\tau\tau$ invariant mass cannot be calculated from the quantities measured by experiments at hadron colliders, as there are at least two neutrinos in the final state. The only available information is the total transverse momentum of all neutrinos, estimated from the missing transverse energy, denoted as $\overline{E}_T^\text{miss}$. Even though the momenta of the remaining neutral and charged decay products of the $\tau$ lepton (visible products, denoted $\tau_\text{vis}$) can be measured precisely, there are two main difficulties:
\begin{enumerate}[(a)]
\item the incomplete description of the final-state kinematics;
\item the $\overline{E}_T^\text{miss}$ resolution, which is one of the least precisely reconstructed observables in a hadronic environment.
\end{enumerate}

Two main approaches to mitigate these issues, adopted by the ATLAS~\cite{Matrix_Mass_Calculator} and CMS~\cite{SVfit} Collaborations, are based on maximum-likelihood parameter estimation, $\mathcal{L}(m_\text{test}\,|\,\text{data})$, where $m_\text{test}$ is the tested $\tau\tau$ invariant mass hypothesis and ``data'' denotes all information available from the detector. However, the two collaborations define the likelihood in different ways.

In the case of the ATLAS Collaboration, the likelihood is based on the probability density functions (PDFs) of the distance between neutrinos and visible $\tau$ decay products~\cite{Matrix_Mass_Calculator}. Hypotheses for the neutrino directions are generated, and the one yielding the maximum likelihood value is chosen. The reconstructed neutrino directions, together with the measured visible momenta, allow for the calculation of the $\tau\tau$ invariant mass. The PDFs are obtained from a large sample of Monte Carlo simulations of $Z \rightarrow \tau\tau$ decays.

The CMS Collaboration adopted a matrix element--based technique in two versions: one using the full matrix element for $pp \rightarrow H \rightarrow \tau\tau \rightarrow X$ with all relevant normalization factors included (denoted Secondary Vertex Fit, SVfit), and a simplified version (classic Secondary Vertex Fit, cSVfit)~\cite{SVfit}. In the simplified version, only terms corresponding to the $H \rightarrow \tau\tau \rightarrow X$ decay are retained. Both methods show similar performance in terms of mass resolution, but the simplified method requires significantly less computational time—around $0.25$~s on a modern computer. However, this is still too slow for complex analyses, where systematic uncertainty estimates require multiple invariant-mass calculations per event and the event samples to be processed are of the order of $10^6$ events.

In this article, we introduce a fast algorithm (fastMTT) inspired by cSVfit. We combine the idea of cSVfit with the well-known collinear approximation (CA)~\cite{Ellis:1987xu}. By applying the CA to the likelihood calculation, we achieve an approximately two orders of magnitude speedup in execution time while maintaining a comparable mass resolution.

The algorithm has been briefly described in the proceedings of the Epiphany Conference 2025 \cite{Matyszkiewicz:2025gal}, where only the final analytical form of the method was presented. The present paper provides a complete and self-contained description of the algorithm, including a detailed motivation and derivation of the underlying formulae, the introduction of a regularization term, and additional features such as an uncertainty estimate and the reconstruction of the $\tau$-lepton kinematics. Moreover, the method's performance is studied in greater detail and compared to existing approaches.

\section{Maximum likelihood estimation}

We estimate the invariant mass of the di-tau system using the matrix element method and maximum likelihood estimation. 
In our simplified approach, we neglect the likelihood components, responsible for the QCD processes in the initial proton-proton scattering. 
We also do not include specific normalization factors because they are not relevant for the position of the likelihood extremum. We model the simplified likelihood using the following formula:

\begin{align} \label{eq:likelihood}
\mathcal{L}(\text{data}|m_\text{test}) = 
& \,\mathcal{N} 
\int d\Phi_n \, 
\left|\mathcal{M}(m_\text{test})\right|^2 W(\text{data}|m_\text{test}),
\end{align}
where $\text{data}$ stands for the observed data, $d\Phi_n$ are the phase space volume elements of both $\tau$ products, $\mathcal{M}(m_\text{test})$ is the matrix element responsible for the investigated process ($H \rightarrow \tau^+\tau^-$), $W(\text{data}|m_\text{test})$ is a transfer function between true and observed values of variables and $\mathcal{N}$ is a normalization factor. In Eq.~\ref{eq:likelihood}, $m_\text{test}$ denotes the tested di-$\tau$ invariant mass hypothesis, for which the likelihood is evaluated.

\subsection{The phase space} \label{phase_space}

In standard $H \rightarrow \tau^+ \tau^-$ analyses, $\tau$ decay channels are usually divided into three distinct categories:

\begin{enumerate}
\item $\tau^- \rightarrow e^- \overline{\nu}_e \nu_\tau$,
\item $\tau^- \rightarrow \mu^- \overline{\nu}_\mu \nu_\tau$,
\item $\tau^- \rightarrow \nu_\tau + \text{hadrons}$, usually denoted as $\tau_h$.

\end{enumerate}

Regardless of the decay channel, among the products we will always find neutrinos that are undetectable by the detector. To make the analysis clearer, we will introduce the notion of visible and invisible four-momentum. 
The $(E^\text{vis}, \overline{p}^\text{vis})$ will refer to the four-vector of electrons, muons, or hadrons in the final states, while $(E^\text{inv}, \overline{p}^\text{inv})$ will describe the sum of neutrinos' four-momenta. We will also denote $x$ as the fraction of energy that is carried by the visible products of the decay:

\begin{equation}x = \frac{E^\text{vis}}{E^\text{vis}+E^\text{inv}} = \frac{E^\text{vis}}{E^\tau}. \label{x_fraction}
\end{equation}

\begin{figure}[H]
    \centering
    \includegraphics[width=0.8\textwidth]{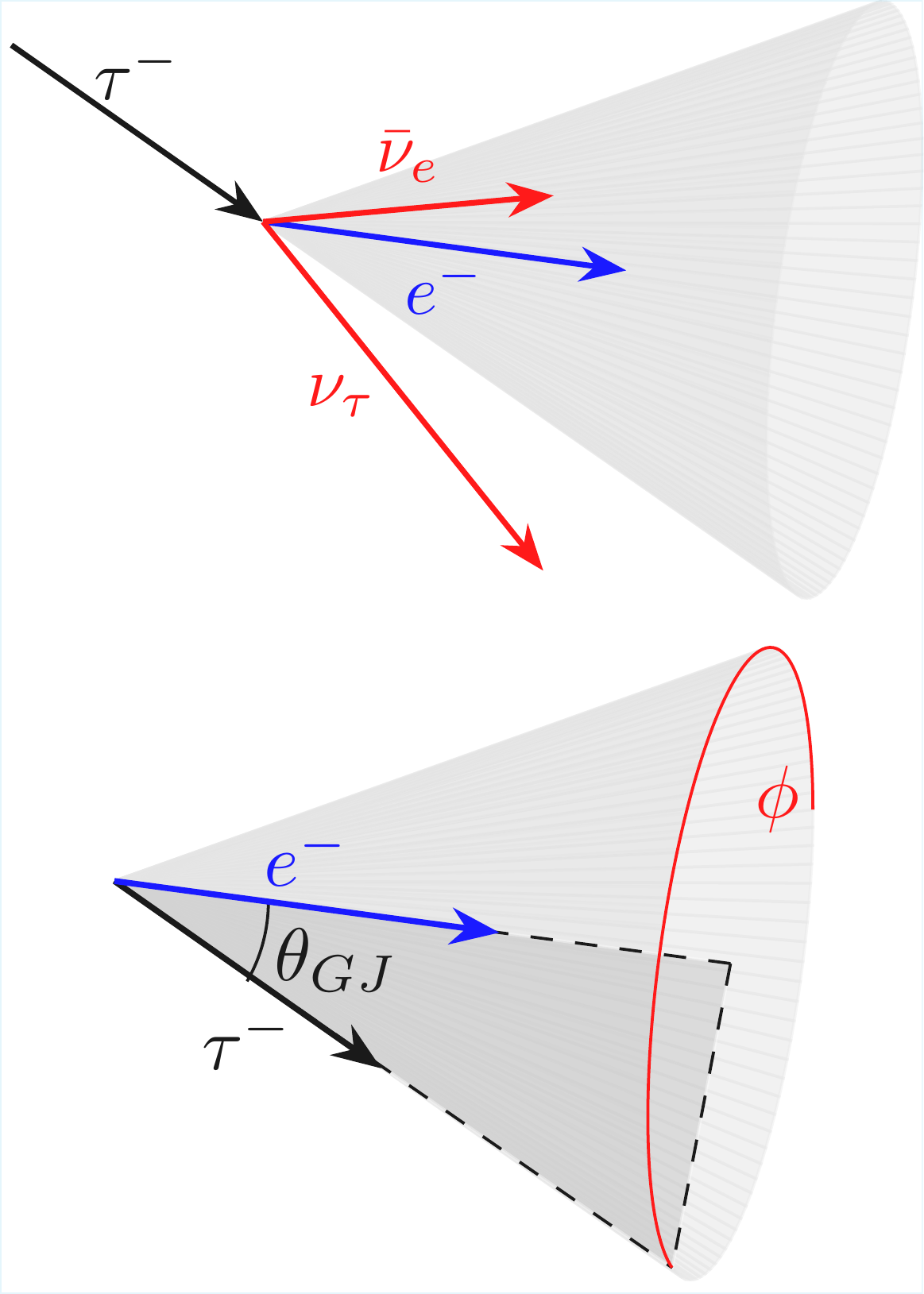}
    \caption{
    Example of tau decay in the electron channel $\tau^- \rightarrow e^- \nu_\tau \bar\nu_e$.
    The top picture shows the division of decay products -- invisible products (neutrinos) are marked with red, while visible products are marked with blue. Light-grey color marks the cone defined by the Gottfried-Jackson angle around $e^-$.
    Bottom picture shows the definition of Gottfried-Jackson $\theta_{GJ}$ and the $\phi$ angles.}
    \label{fig:GJ_angle}
\end{figure}

To fully describe the kinematics of a decaying tau, we need three additional parameters that describe the kinematics of invisible products. It is customary to use $\theta_\text{GJ}$ -- Gottfried-Jackson angle \cite{Gottfried-Jackson} \cite{Tau_GJ}, which is the angle between the tau lepton momentum and sum of visible momenta in the tau rest frame; $\phi$ -- azimuth angle describing the position of the $\tau$ momentum on a cone centered around the visible momentum (Fig.~\ref{fig:GJ_angle}) and $m_{\nu \nu}$, an invariant mass of the neutrino system. It is worth noting that in the case of hadronic decay, $m_{\nu \nu}$ is equal to zero (assuming $m_\nu = 0$), as there is only one neutrino in the final state.

Since taus originate from a heavy particle ($m \gg m_\tau$) decay, it is therefore reasonable to assume that the tau decay products in the laboratory frame are collinear (due to the Lorentz boost) with the direction of the tau lepton momentum.
For light decay products, the maximal value of $\theta_{GJ}$ can be approximated as:

\begin{equation}
    \theta_{GJ} \sim \frac{1}{\gamma},
\end{equation}

where $\gamma$ is the Lorentz factor of the $\tau$ lepton in the laboratory frame. For example, assuming that the energy from the Higgs boson decay is split equally between two $\tau$ leptons, we have $\gamma \approx 35$, and $\theta_{GJ}$ is around $1.63^\circ$. 
Therefore, we assume

\begin{equation}
    \theta_\text{GJ} = 0
\end{equation}
and leave the $\phi$ angle degenerate as a result. This translates to a physical interpretation in which both the visible and invisible systems are collinear with the direction of the $\tau$ lepton momentum, and:

\begin{equation}
    \overline{p}_\text{inv} = \overline{p}_\text{vis} \frac{1 - x}{x}.
\end{equation}

As a consequence, we are left with a significantly reduced phase space. When calculating the likelihood, we typically need to integrate over the four-dimensional phase space $d\Phi$ of each $\tau$ lepton. However, in this case, we are left with only a one- or two-dimensional phase space (depending on the $\tau$ decay channel) for each $\tau$ lepton:

\begin{equation}
d\Phi_n =
\begin{cases}
dx \, dm_{\nu\nu}, & \text{for leptonic decay}, \\
dx, & \text{for hadronic decay}.
\end{cases}
\end{equation}

The range of the integration variables is constrained by the kinematics of tau lepton decay. For $x$ it is:

\begin{equation}
    \left(\frac{m_\text{vis}}{m_\tau}\right)^2 < x < 1.
\end{equation}

While for $m_{\nu\nu}$ we get:
\begin{equation}
    0 < m^2_{\nu\nu} < (1-x)m_\tau^2.
\end{equation}

Detailed derivation of those limits can be found in the \ref{appendix: kinematical limits}.

Finally, with the collinear approximation, we obtain a simple formula for the $\tau \tau$ invariant mass as a function of the invariant mass of the visible products, and the $x$ parameters:
\begin{equation}
m_{\tau\tau} = \frac{m_\text{vis}}{\sqrt{x_1 x_2}}, \label{m_tau_tau}
\end{equation}
where $x_1$ and $x_2$ are the fractions of energy carried by the visible products of each $\tau$, respectively, and $m_\text{vis}$ is the visible mass of the full di-$\tau$ system.

In the following, the likelihood is evaluated as a function of the parameters $(x1, x2)$, with the test mass $m_\text{test}$ understood as the derived quantity defined in Eq.~\eqref{m_tau_tau}.

\subsection{Matrix elements} \label{Matrix_elements}

The matrix element is maximally simplified to include only the Breit–Wigner
distribution of the $\tau$ lepton:

\begin{equation}
\left|\mathcal{M}(p, m_\text{test})\right|^{2}
= \prod_{i=1}^{2} \left|\mathrm{BW}^{(i)}_\tau\right|^{2},
\end{equation}

where $p$ denotes tau four-momentum.

The Breit–Wigner distribution is substituted using the narrow-width approximation:

\begin{equation}
\left|\mathrm{BW}^{(i)}_\tau\right|^{2}
= \frac{\pi}{m_\tau \Gamma_\tau}\,
\delta\!\left(m_\tau^2 - p^2_\text{test}\right).
\end{equation}

After integrating over the Dirac delta distribution, we are left solely with the
phase space described in Section~\ref{phase_space}. The likelihood is then given
by the phase-space volume consistent with the observed data and with the values
of $x_{1,2}$ under the test mass:

\begin{equation}
\begin{split}
\mathcal{L}(\text{data}|x_1, x_2) \sim
\int d\Phi_n \,
\left|\mathcal{M}(m_\text{test} | x_1, x_2)\right|^2 W(\text{data}|x_1, x_2) = &\\
\int d\Phi_n \, \delta\left(m_\text{test} - \frac{m_\text{vis}}{\sqrt{x_1 x_2}}\right)
W(\text{data}|x_1, x_2)
\end{split}
\end{equation}

In the likelihood calculation, we omit the normalization factor, as it is the
same for all events and therefore does not affect the position of the likelihood
maximum.

\subsection{Transfer functions}

We assume the $\overline{E}_T^\text{miss}$ uncertainty is much bigger than the uncertainty of the kinematic variables of the visible products of the $\tau$ decay.
Therefore, we approximate the transfer functions for the visible products with identity. The $\overline{E}_T^\text{miss}$ transfer function is modeled with the normal distribution, with event-by-event covariance matrix as it was in the original cSVfit algorithm \cite{SVfit}.
The covariance matrix of the reconstructed $\overline{E}_T^\text{miss}$ is used as an input of the algorithm. If such an event-by-event covariance matrix is not available, a constant covariance matrix can be used.

\begin{equation}
    W(\overline{E}_T^\text{rec}|\overline{E}_T^\text{test}) = 
    \mathcal{N} \exp\left(-\frac{1}{2}(\overline{E}_T^\text{test} - 
    \overline{E}^\text{rec}_T)^{T} {V^{-1}_\text{MET}}(\overline{E}_T^\text{test} - 
    \overline{E}^\text{rec}_T)\right), \label{TF_MET}
\end{equation}

where $(\overline{E}_T^\text{test} - \overline{E}^\text{rec}_T)$ is the difference between the test and reconstructed missing transverse energy, and $V_\text{MET}$ is the reconstruction covariance matrix. The $\overline{E}_T^\text{test}$ is calculated as the sum of the momenta of the invisible $\tau$ decay products, calculated from the $x_1$ and $x_2$ hypotheses and collinear approximation:
\begin{equation}
\overline{E}_T^\text{test} = \frac{\overline{p}^\text{vis}_1(1-x_1)}{x_1} + \frac{\overline{p}^\text{vis}_2(1 - x_2)}{x_2}.\label{eq:MET_from_x1x2}
\end{equation}

\subsection{Final likelihood} \label{subsection: likelihood separation}

For the final step, we factorize the likelihood by taking $\overline{E}_T^\text{test}$ transfer function outside the integral:

\begin{equation}
    \begin{split}
    \mathcal{L}(\text{data}|x_1, x_2) \sim
    W(\text{data}|x_1, x_2)
    \int d\Phi_n \,
    \left|\mathcal{M}(m_\text{test} | x_1, x_2; m_\text{vis})\right|^2.
    \end{split}
\end{equation}

This factorization is motivated by two observations. First, each component independently peaks near the true invariant mass of the system; the corresponding validation studies are shown in \ref{appendix: likelihood separation}. Second, once separated, the transfer function and the phase–space term capture largely distinct pieces of information: the transfer function constrains the true missing transverse energy primarily, while the phase–space factor uses all remaining kinematic information of the event. Consequently, the residual correlation between them is weak.

Given two likelihood contributions that independently favor the true mass, using their product provides a practical and robust estimator of the overall likelihood profile.

With this factorization, the phase–space integral becomes analytically tractable, which is essential for achieving the computational performance required by the algorithm.

The form of the integrand depends on the $\tau\tau$ decay channel. In the hadronic decay channel, where the integration over the neutrino invariant mass is not performed, we obtain:

\begin{equation}
    \begin{split}
    \int d\Phi_n \,
    |\mathcal{M}_{h l}|^{2} = & \\
    \int_{x_{1,\text{min}}}^1 dx_1 \int_{x_{2,\text{min}}}^1 dx_2 \, 
    \delta\left(m_\text{test} - \frac{m_\text{vis}}{\sqrt{x_1 x_2}}\right) = &\\
    \int_{x_\text{min}}^{x_\text{max}} dx_2 \, \frac{2m^2_\text{vis}}{m^3_\text{test}} \frac{1}{x_2} = & \\
    = \frac{2m^2_\text{vis}}{m^3_\text{test}} 
    \log \left(\frac{x_\text{max}}{x_\text{min}}\right)&,
    \end{split}
    \label{eq:int_hh}
    \end{equation}
    
where $x_\text{min} = \max\left( x_{2, min}, \left(\frac{m_\text{vis}}{m_\text{test}}\right )^2\right)$ and $x_\text{max} = \min\left( 1, \left(\frac{m_\text{vis}}{m_\text{test}}\right )^2\frac{1}{x_{1, min}}\right)$.
For the semi-leptonic decay, we need also to consider integration over $m_{\nu \nu}$ 
for the leptonic $\tau$ decay:
    
\begin{align}
    \int d\Phi_n \,|\mathcal{M}_{h l}|^{2}
    &= \int_{x_{1,\text{min}}}^1 dx_1 \int_{x_{2,\text{min}}}^1 dx_2 \int_0^{(1 - x_1)m_\tau^2} dm^2_{\nu \nu} \, 
    \delta\left(m_\text{test} - \frac{m_\text{vis}}{\sqrt{x_1 x_2}}\right) \nonumber \\
    &= m_\tau^2\frac{2m^2_\text{vis}}{m^3_\text{test}} \left(\log \left(\frac{x_\text{max}}{x_\text{min}}\right) + \left(\frac{m_\text{vis}}{m_\text{test}}\right)^2 \left( \frac{1}{x_\text{max}} - \frac{1}{x_\text{min}}\right)\right).
    \label{eq:int_hl}
\end{align}
    
Finally, for the di-leptonic decay, we need to integrate over the $m_{\nu \nu}$ for both tau leptons:
    
\begin{align}
    \int d\Phi_n \,|\mathcal{M}_{l l}|^{2}
    &= m_\tau^4 \frac{2m^2_\text{vis}}{m^3_\text{test}} \Bigg( 
    \left(1 + \left(\frac{m_\text{vis}}{m_\text{test}}\right)^2\right) \log \left(\frac{x_\text{max}}{x_\text{min}}\right) \nonumber \\
    &\quad + \left(\frac{m_\text{vis}}{m_\text{test}}\right)^2 \left(\frac{1}{x_\text{max}} - \frac{1}{x_\text{min}}\right) 
    - \left(x_{2,\text{max}} - x_{2,\text{min}}\right) \Bigg).
    \label{eq:int_ll}
\end{align}

The $x_{1,2,\text{min}}$ depend on the decay mode and can be found in the \ref{appendix: kinematical limits}.

At the end, we obtain the following likelihood:

\begin{align}
\mathcal{L}(\text{data}|x_1, x_2) = 
W(\overline{E}_T^\text{rec}|\overline{E}_T^\text{test}) I
\label{eq:final_likelihood}
\end{align}

with integral $I$ given by equations~\ref{eq:int_hh} - \ref{eq:int_ll}.

\subsection{Regularization}

Following ~\cite{SVfit}, we also consider a penalized maximum-likelihood approach, introduced to suppress the long tails of the reconstructed di-$\tau$ invariant-mass distribution. The motivation for introducing this penalty is to reduce the extended high-mass tails originating from reconstructed $Z^0$ boson events, which would otherwise degrade the sensitivity of searches for heavy
neutral resonances. This effect is particularly important because the $Z^0$ boson has a much larger production cross section and therefore constitutes a dominant background that must be properly accounted for. To this end, Ref.~\cite{SVfit} introduces an additional penalty term in the likelihood:

\begin{equation}
\mathcal{L^{\text{reg}}}(\text{data}|x_1, x_2) = \mathcal{L}(\text{data}|x_1, x_2) + r(\Theta).
\end{equation}

In our case, we have used a different approach. The likelihood shape is modified in the following way:

\begin{itemize}
\item test-mass scaling: 
\begin{equation}m_\text{test} \rightarrow \alpha m_\text{test},\end{equation}
\item stronger suppression of the likelihood for the 
      large values of the test mass with the following substitution
\begin{equation}\frac{m^2_\text{vis}}{m^3_\text{test}} \rightarrow 
    \frac{m^2_\text{vis}}{m^\beta_\text{test}}, \label{eq:beta} \end{equation}
\end{itemize}

in the Jacobi factor in equations~\ref{eq:int_hh} - \ref{eq:int_ll}.

The parameter $\alpha$ modifies the peak position of the reconstruction, and $\beta$ modifies both the peak and the width of the reconstructed masses.

Both parameters were chosen using a trial-and-error procedure to achieve a performance comparable to that of "cSVfit". The parameter $\alpha$ was fixed to $\alpha = 1/1.1$, while $\beta$ was chosen depending on the decay channel: $\beta = 6$ for fully hadronic decays of two $\tau$ leptons, $\beta = 2$ for semi-leptonic decays (one leptonic and one hadronic $\tau$ decay), and $\beta = 3.5$ for fully leptonic decays. The comparison is performed using simulated $Z^0 \to \tau\tau$ events and signal samples of $H \to \tau\tau$, as described in Section~\ref{sec: reco_resolution}.

We note, however, that the use of constant factors is likely not optimal. The values of $\alpha$ and $\beta$ could be further optimized for specific experimental workflows, in particular for different $\overline{E}_T^{\text{miss}}$ reconstruction techniques.

\subsection{Likelihood maximization}

Instead of scanning the likelihood in the $m_{\tau \tau}$ space, we perform the scan in the ($x_1, x_2)$ space, where each point determines  $m_{\tau\tau}$ as for Eq.~\ref{m_tau_tau}. Points with the same value $m_{\tau \tau}$ differ by $\overline{E}_T^\text{test}$ hypothesis according to Eq. ~\ref{eq:MET_from_x1x2}.

We use the grid search method and evaluate the above formula on a parameter grid of $x_1$ and $x_2$. We use a grid with a resolution of $100 \times 100$. We select the $x_1$ and $x_2$ values that yield the highest likelihood and use them to compute the momenta of taus (eq. \ref{x_fraction}) and then the invariant mass of the system:

\begin{equation}
m^2_\text{test} = \left( \frac{E^\text{vis}_1}{x_1}+\frac{E^\text{vis}_2}{x_2}\right)^2 - \left( \frac{p^\text{vis}_1}{x_1}+\frac{p^\text{vis}_2}{x_2}\right)^2 = \frac{m^2_\text{vis}}{x_1x_2}.\label{mass from x-es}
\end{equation}

\begin{figure}[H]
     \centering
     \includegraphics[width=\textwidth]{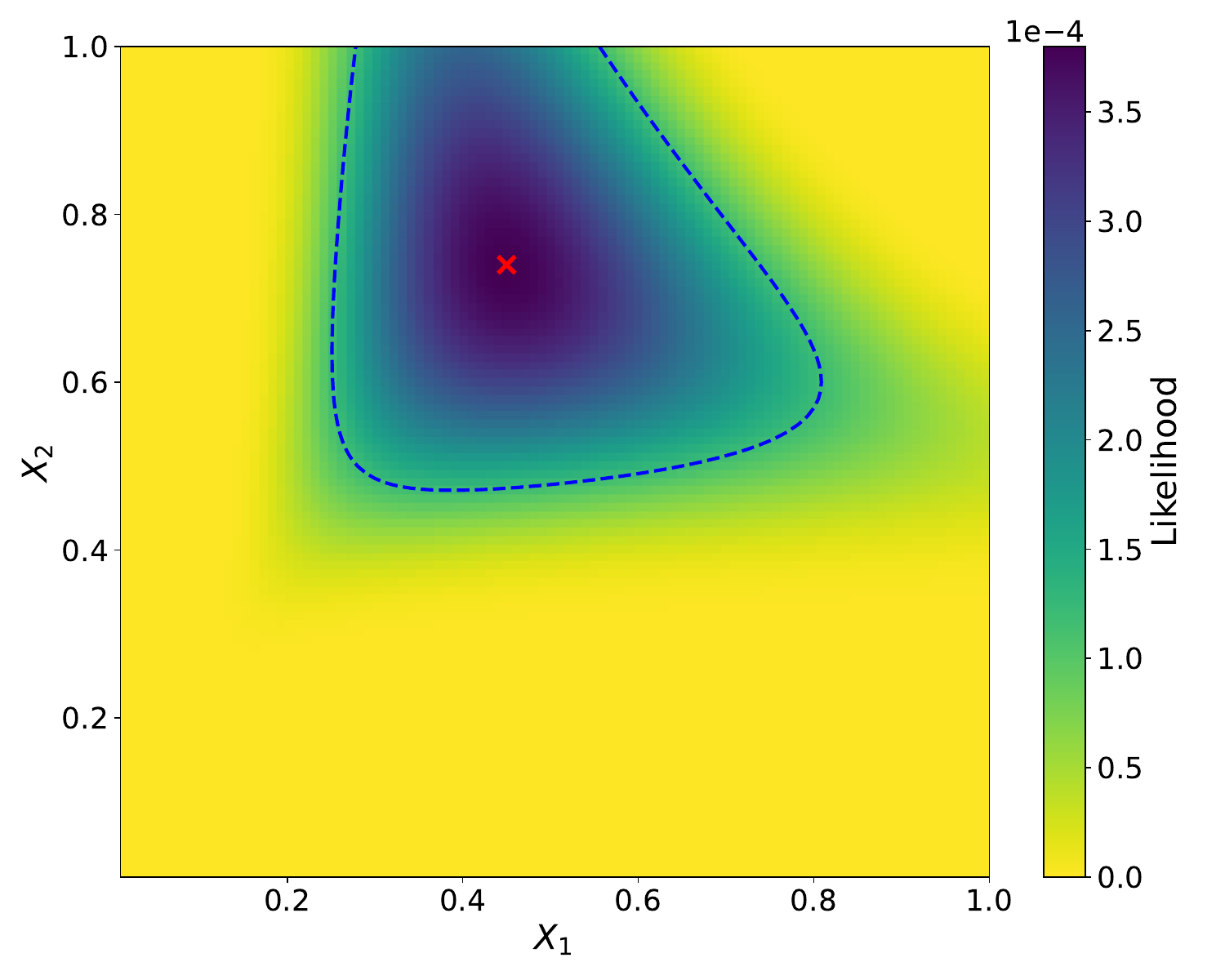}
     \caption{
     The likelihood map on the ($x_1, x_2$) plane for a single event. The red cross marks the position of the likelihood maximum, while the blue contour limits the area in which there is $\sim 68\%$ (1 $\sigma$) probability to choose the true mass.}
     \label{fig:likelihood}
     \includegraphics[width=\textwidth]{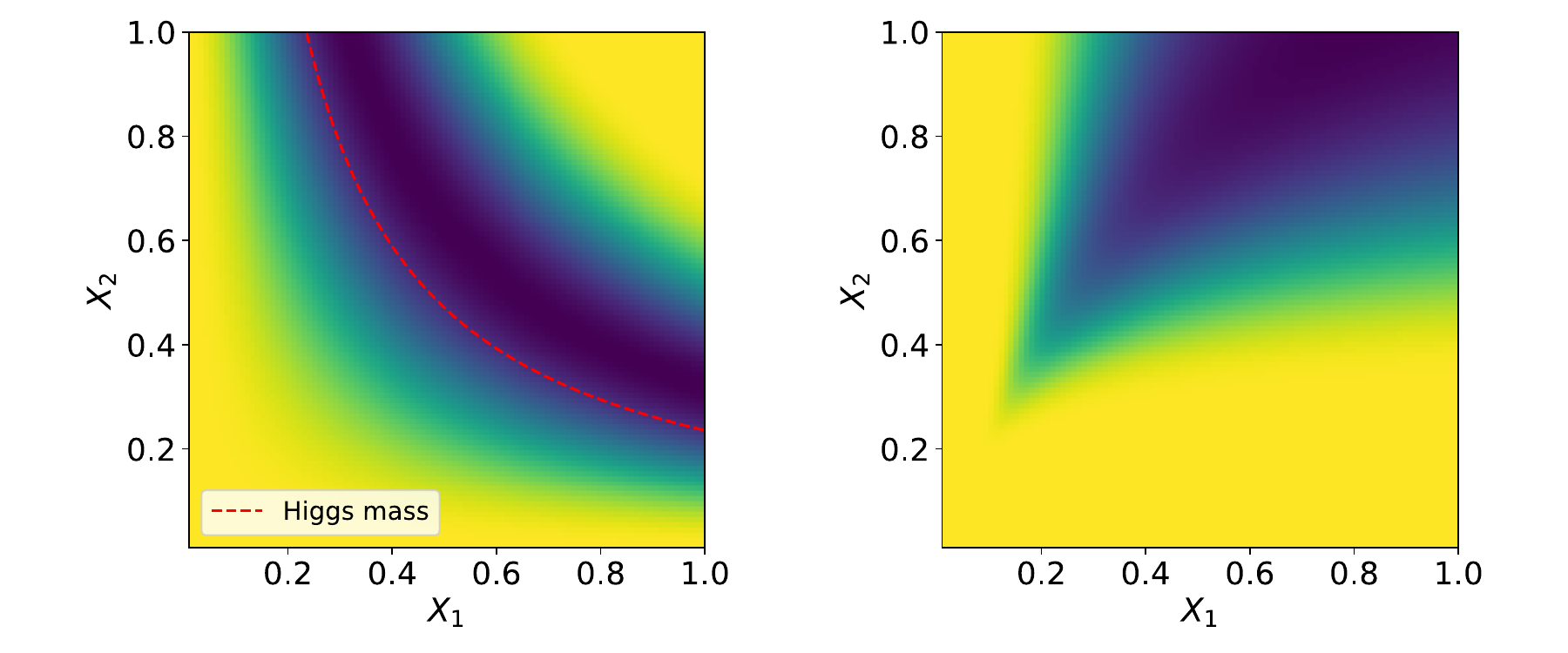}
     \caption{
     Comparison of two main components of the final likelihood: integral over the phase space (left) and transfer function (right). The red line marks curve along with $m_{\text{test}}$ = 125~GeV/c$^{2}$.}
     \label{fig:likelihood_subcomponents}
 \end{figure}

An example of the likelihood scan is shown in Figure \ref{fig:likelihood}. 
Fig. \ref{fig:likelihood_subcomponents} contains a comparison of two main components of the likelihood: the integral of the phase space and the transfer function.

\subsection{Uncertainty estimation} \label{error_chapter}

The grid scan of the likelihood provides full map of the likelihood as a function of the $x_1$ and $x_2$ parameters. This allows to estimate the uncertainty of the mass reconstruction in a time-efficient way using the likelihood profile method. We make use of the Wilks~\cite{Wilks:1938dza} 
theorem, which implies that:

\begin{equation}
\mathcal{L}(\theta) = \mathcal{L}(\hat\theta)\exp(-\frac{\chi^2}{2}),
\label{eq:Wilks}
\end{equation}
where $\theta$ is the reconstructed variable, $\hat\theta$ is its maximum-likelihood estimate, $\log \mathcal{L}(\hat\theta)$ is the maximum likelihood, and $\chi^2$ is the value from the $\chi^2$ distribution for 2 degrees of freedom. For example for confidence level of $\sim 68\%$ (which in normal distribution is equivalent to $1 \sigma$), $\chi^2$ is $2.3$ and for the confidence level $99.7\%$ ($3\sigma$) it is $9.2$.
An advantage of this method is that it depends only on the value of the maximum of the likelihood. Therefore, it can be used without calculating the normalization factor. We use the likelihood maps to find the contour of the region with the masses of the assumed confidence level. 
An example of contour is shown in Fig. \ref{fig:likelihood}. The $x_{1,2}$ values are translated to the mass values using eq. \ref{mass from x-es}. The confidence level is spanned by the minimal and maximal values of the mass.

As the derived confidence levels correspond to the simplified likelihood (following Sec.~\ref{subsection: likelihood separation}), we expect that they may be either over- or underestimated. Therefore, we propose calibrating them using the following procedure.

First, the likelihood is evaluated using simulated Monte Carlo samples of
Higgs and $Z^0$ boson events. Next, the distribution and variance of the
normalized residual (pull) are computed:

\begin{equation}
z = \frac{m_\text{reco} - m_\text{true}}{\sigma_\text{reco}},
\label{pull}
\end{equation}
where $m_\text{reco}$ denotes the reconstructed mass, $m_\text{true}$ the true
mass, and $\sigma_\text{reco}$ the estimated uncertainty of the reconstructed
mass.

The quantity $z$ should asymptotically follow a normal distribution~\cite{MLE-uncertainties} $\mathcal{N}(\mu_z, \sigma_z)$. Therefore, the reconstructed uncertainty is adjusted according to

\begin{equation} \label{pull calibration}
\sigma_\text{calibrated} = \sigma_\text{reco}\,\sigma_z.
\end{equation}

\subsection{Additional mass constraint} \label{Constraint}

The likelihood formulation enables the straightforward incorporation of additional constraints, thereby extending its application from background discrimination to energy reconstruction.

Suppose the analyzed events are expected to originate from a known resonance, such as the $Z^0$ or the Higgs boson. In that case, one can add a mass constraint to the likelihood to improve the $x_{1,2}$ resolution and therefore the individual $\tau$ energies. The directions cannot be constrained, as the algorithm assumes that the taus are collinear with measured visible products.

The mass constraint is incorporated by multiplying the likelihood function by a new component, $C(m)$, which modifies the main likelihood:

\begin{equation}
\mathcal{L}(\text{data}|x_1, x_2) = 
 \mathcal{L}_b(\text{data}|x_1, x_2)\, C(m),
\end{equation}
where $\mathcal{L}_b(\text{data}|m_\text{test})$ represents the baseline likelihood, as defined in Eq. \ref{eq:final_likelihood} and $m$ is a mass of selected resonance (e.g. $125$ GeV for Higgs).

We model the constraint using a normal distribution:

\begin{equation}
C(m) = \exp\left(-\frac{(m_\text{test} - m)^2}{2\sigma^2}\right),
\end{equation}
where $\sigma$ is chosen arbitrarily. When selecting $\sigma$, one must balance between a constraint that is too weak, rendering its effect negligible, and one that is too strong, turning the reconstruction algorithm into a pure discriminator. We have evaluated the performance of the algorithm using the transverse momentum ($p_T$) of the reconstructed taus as a figure of merit. 
This choice is motivated by the fact that information about the $p_T$ of individual tau leptons is important in $H\rightarrow\tau^+\tau^-$ analyses, such as CP phase measurements~\cite{Higgs_CP}.

For the Higgs boson, the best performance was achieved with $\sigma = 7$ GeV. However, the results remained largely stable as long as $\sigma$ was kept within the range of $2-15$ GeV.

The mass constraint effectively acts as a Bayesian prior on the di-$\tau$ invariant mass, encoding prior knowledge of the resonance mass. While this improves the precision of the reconstructed $\tau$-lepton energies, it simultaneously biases the reconstructed mass toward the Higgs mass hypothesis. As a result, the likelihood no longer serves as an effective discriminator between signal and background.

\begin{figure}[h]
    \centering
    \includegraphics[width=0.8\textwidth]{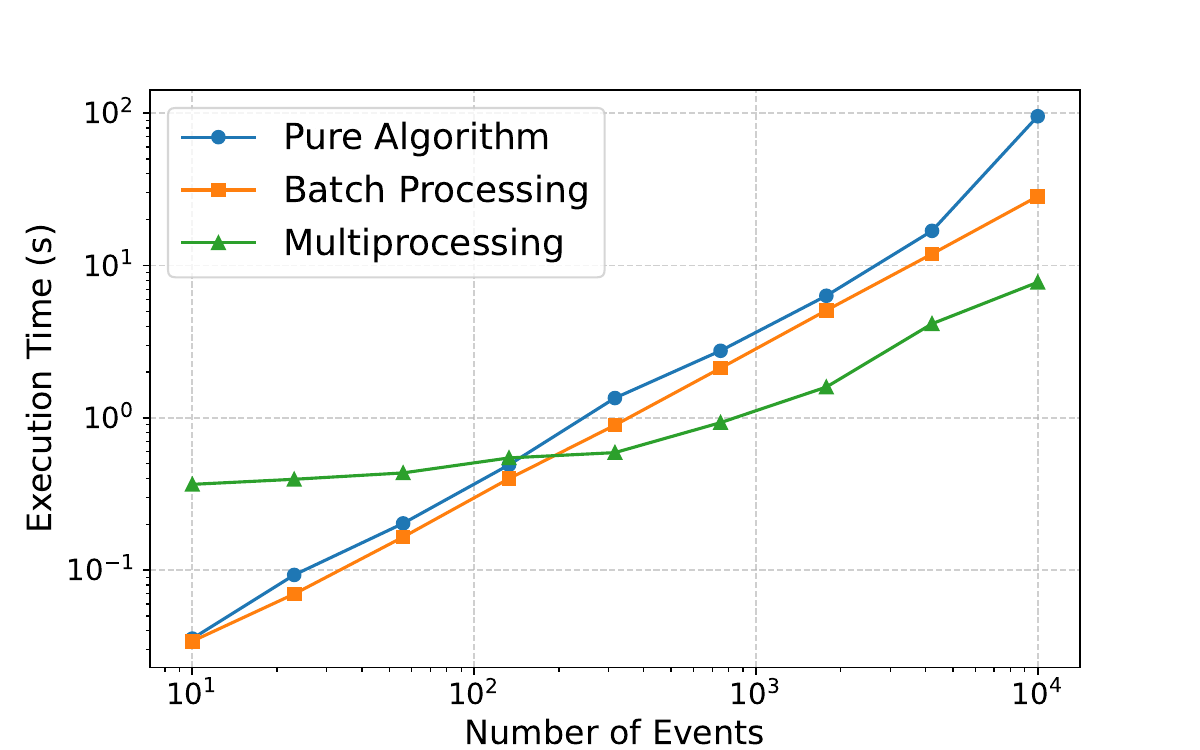}
    \caption{Time needed to calculate the masses for a given number of cases. The figure shows a comparison between the NumPy implementation (blue line), using batch processing (orange line), and both batch and multiprocessing (green line). All computations were performed locally on a system equipped with two AMD Opteron 6320 processors (16 logical cores, x86\_64 architecture, 2.8 GHz).}
    \label{fig:time performance}
\end{figure}

\begin{figure}[h]
    \centering
    \includegraphics[width=\textwidth]{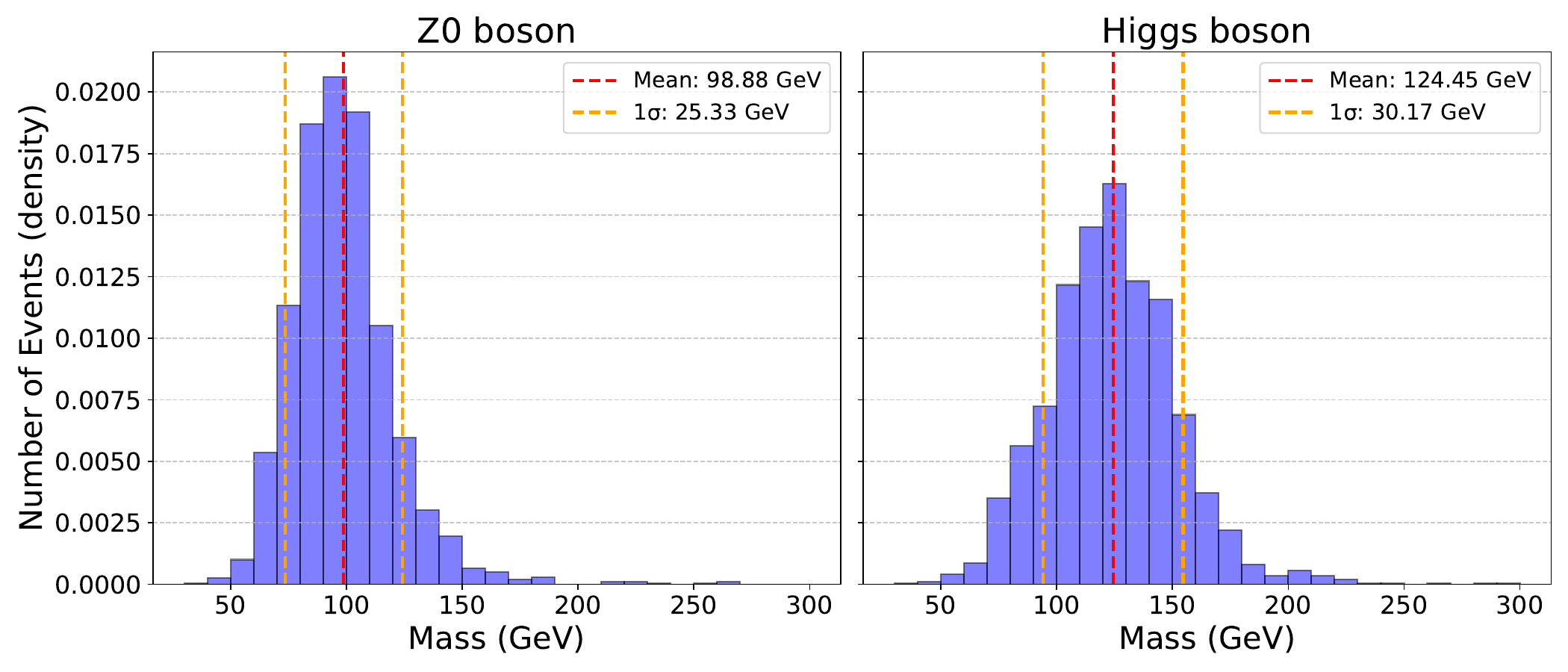}
    \caption{Reconstruction of the $Z^0$ (left) and Higgs (right) bosons' mass with the fastMTT algorithm. The x-axis represents the invariant mass of the reconstructed di-tau system, while the y-axis represents event density. For the Higgs boson, the fractions of fully hadronic, semi-leptonic, and fully leptonic events are $61\%$, $34\%$, and $5\%$, respectively, while for the $Z^0$ boson they are $65\%$, $30\%$, and $5\%$.}
    \label{fig:Higgs and Z0}
\end{figure}

\begin{table}[h]
\centering
\caption{Summary of mass reconstruction performance for different decay channels. Mean and standard deviation are reported for each decay channel.
Had-had denotes the fully hadronic decay channel, had-lep the semi-leptonic channel, and lep-lep the fully leptonic channel. The parameter $\beta$ represents the regularization strength defined in Eq.~\ref{eq:beta} and is chosen independently for each decay channel.}
\label{table:mass_in_different_DMs}
\begin{tabular}{|c|ccc|}
\hline
Process & Mass mean [GeV] & Mass std. dev. [GeV] & $\beta$ \\
\hline
Higgs (had-had) & $124.5$ GeV & $29.7$ GeV & $6$ \\
Higgs (had-lep) & $124.6$ GeV & $32.3$ GeV & $2$ \\
Higgs (lep-lep) & $124.7$ GeV & $29.1$ GeV & $3.5$ \\
\hline
$Z^0$ (had-had) & $99.7$ GeV & $23.8$ GeV & $6$ \\
$Z^0$ (had-lep) & $98.2$ GeV & $28.5$ GeV & $2$ \\
$Z^0$ (lep-lep) & $97.0$ GeV & $26.7$ GeV & $3.5$ \\
\hline
\end{tabular}
\end{table}

\begin{figure}[h]
    \centering
    \includegraphics[width=\textwidth]{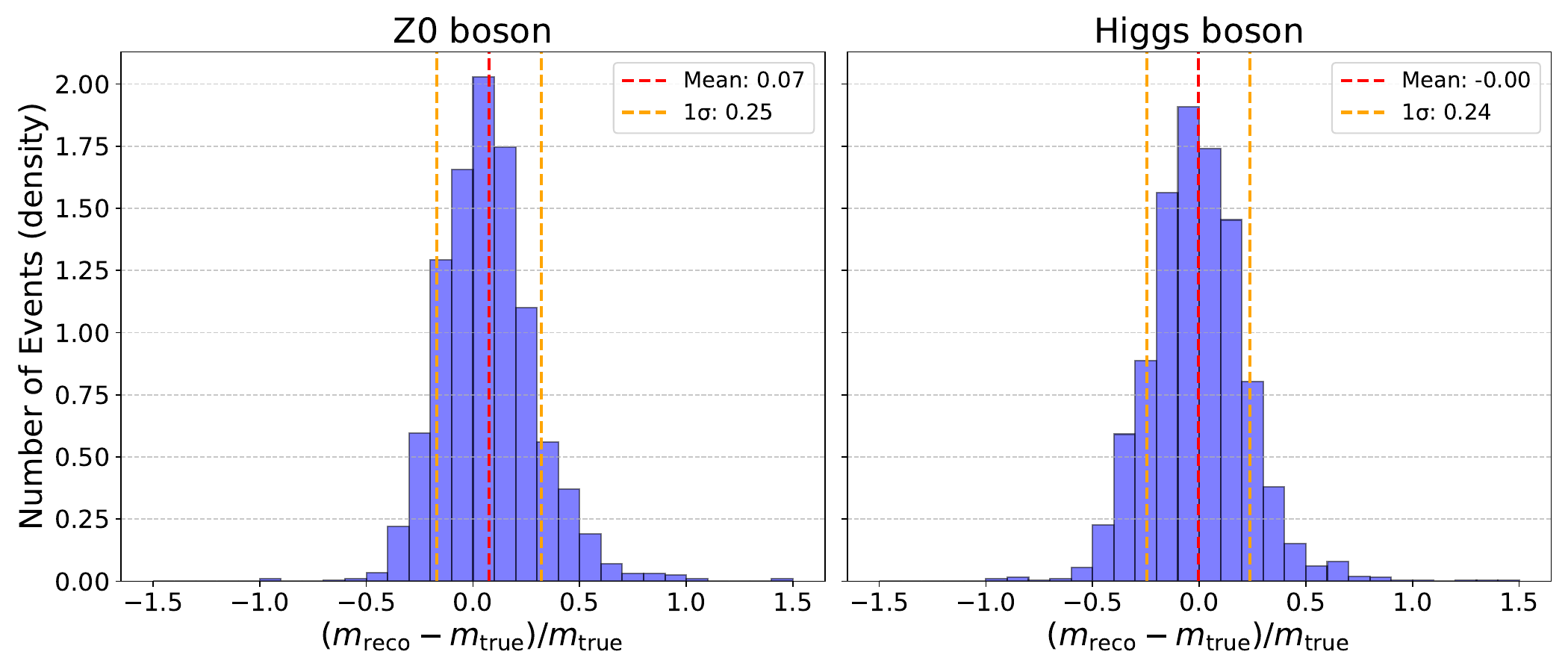}
    \caption{Relative resolution of $\tau\tau$ reconstruction for $Z^0$ (left) and Higgs (right) bosons. The x-axis represents the relative resolution of the mass reconstruction, while the y-axis represents event density. For the Higgs boson, the fractions of fully hadronic, semi-leptonic, and fully leptonic events are $61\%$, $34\%$, and $5\%$, respectively, while for the $Z^0$ boson they are $65\%$, $30\%$, and $5\%$.}
    \label{fig:mass resolution}
\end{figure}

\begin{table}[h]
\centering
\caption{Summary of the reconstruction resolution, defined as $(m_\text{reco} - m_\text{true}) / m_\text{true}$, for different decay channels.
Mean and standard deviation are reported for each decay channel. Had-had denotes the fully hadronic decay channel, had-lep the semi-leptonic channel, and lep-lep the fully leptonic channel. The parameter $\beta$ represents the regularization strength defined in Eq.~\ref{eq:beta} and is chosen independently for each decay channel.}
\label{table:resolution_in_different_DMs}
\begin{tabular}{|c|ccc|}
\hline
Process & Resolution mean & Resolution std. dev. & $\beta$ \\
\hline
Higgs (had-had) & $-0.003$ & $0.24$ & $6$ \\
Higgs (had-lep) & $-0.003$ & $0.26$ & $2$ \\
Higgs (lep-lep) & $-0.003$ & $0.24$ & $3.5$ \\
\hline
$Z^0$ (had-had) & $0.09$ & $0.23$ & $6$ \\
$Z^0$ (had-lep) & $0.06$ & $0.27$ & $2$ \\
$Z^0$ (lep-lep) & $0.06$ & $0.26$ & $3.5$ \\
\hline
\end{tabular}
\end{table}

\section{Performance}

\subsection{Implementation and time efficiency}

The algorithm has been implemented in the Python programming language, and is available from a public repository~\cite{Github}. To optimize performance, we avoid explicit loops and process events simultaneously using NumPy \cite{Numpy} arrays, leveraging the power of vectorization and broadcasting for better execution speed. 
Additionally, we introduce batch processing to mitigate memory allocation issues, which further accelerates the algorithm by requiring class initialization only once. Moreover, we adapt the code for multiprocessing to fully exploit the algorithm’s capabilities.

All computations were performed locally on a system equipped with two AMD Opteron 6320 processors (16 logical cores, x86\_64 architecture, 2.8 GHz). Parallel processing was used via Python’s multiprocessing using a pool with up to 8 worker processes. The time performance of the algorithm is shown in Fig. \ref{fig:time performance}.

Our study shows that using only the fastMTT algorithm, one can process ten thousand events in $46$ seconds, which corresponds to a time budget of $4.6 \, \mathrm{ms}$ per event.

Utilizing batch processing -- which is necessary for handling a large number of events due to the vectorized implementation of the algorithm -- provides an additional speed-up, allowing the same number of events to be processed in approximately $28\,\mathrm{s}$ ($2.8\,\mathrm{ms}$ per event).
Further acceleration is achieved by employing multiprocessing with 8 workers, reducing the total computation time to $7\,\mathrm{s}$, corresponding to $0.7\,\mathrm{ms}$ per event, which is sufficient for the practical needs of physics analyses.

The failure rate of the algorithm is negligible.

\subsection{Reconstruction resolution} \label{sec: reco_resolution}

We have tested the algorithm on Monte Carlo simulated samples of Higgs and $Z^0$ bosons. The datasets used in this study are available in the same public repository~\cite{Github}.

The events were generated with Pythia 8.2 \cite{Pythia} and the CUETP8M1 tune \cite{CMS_tunes} of NNPDF2.3LO parton distribution function at the energy of 14 TeV proton-proton collisions. We used Delphes 3.5.0 \cite{Delphes} with the Delphes\_Card\_CMS.tcl (with $\Delta R$ changed to 0.4) to simulate the response of the detector. We did not include any pile-up effects.

The covariance matrix used by the transfer function was obtained from the Gaussian fit to the differences between the true and reconstructed missing transverse energy $\overline{E}_T^\text{miss}$. We have used a global covariance for all events. It is worth pointing out that for better behavior of the algorithm, one should calculate $\overline{E}_T^\text{miss}$ covariance event-by-event.

The algorithm was tested in its power to differentiate between signal and background (without mass penalty term). The distribution of reconstructed mass is shown in Fig.~\ref{fig:Higgs and Z0}, while the pulls: $(m_\text{reco} - m_\text{true})/{m_\text{true}}$ are shown in Fig. \ref{fig:mass resolution}. Results are also reported for different decay channels in Table~\ref{table:mass_in_different_DMs} for reconstructed masses and Table~\ref{table:resolution_in_different_DMs} for resolution of the reconstruction.

The mass resolutions, defined as the standard deviations of the reconstructed mass distributions, are $25$ GeV and $30$ GeV for the $Z^0$ and Higgs bosons, respectively. The corresponding relative resolutions are $0.25$ for the $Z^0$ boson and $0.24$ for the Higgs.

\begin{figure}[H]
    \centering
    \includegraphics[width=1.05\textwidth]{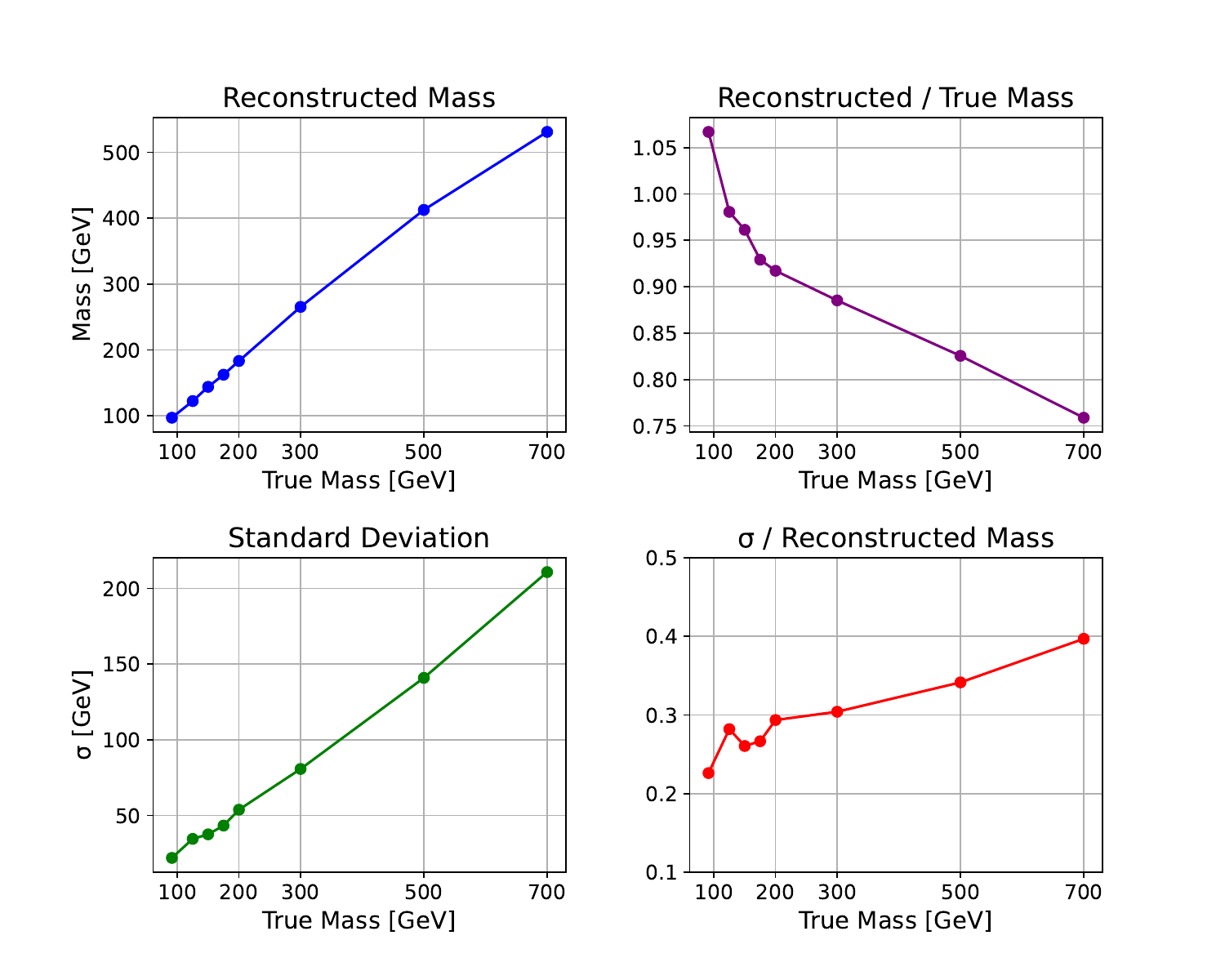}
    \caption{FastMTT performance for hypothetical neutral resonances with masses: $150, 175, 200, 300, 500, 700$ GeV. The figure shows reconstructed mass as a function of true mass (upper left), ratio of the reconstructed and true masses (upper right), reconstructed mass standard deviation (lower left), and ratio of standard deviation and the true mass (lower right).}
    \label{fig:different masses}
\end{figure}

In order to test whether the algorithm is feasible for searches of new neutral resonances, we have evaluated the algorithm for $\tau\tau$ resonances with masses of $150, 175, 200, 300, 500, 700$ GeV. We calculated the mean and standard deviation for the testing sample of each mass and plotted them in Figure \ref{fig:different masses}. 
In this case, one can see that the algorithm is capable of correctly reconstructing masses for low-mass resonances (up to masses of $300$ GeV). In the case of higher masses, the reconstruction of mass is becoming worse; however, it still provides an estimator with a linear response to the true resonance mass.

Finally, we tested the mass–uncertainty estimation by analyzing the normalized residual $z$ defined in Eq.~\ref{pull}. The study was performed on the combined Higgs–$Z^0$ sample, constructed from equal-sized subsamples and separated into three decay channels: fully hadronic, semi-leptonic (with one tau decaying hadronically and the other leptonically), and fully leptonic.

For fully hadronic decays, we obtained a standard deviation of $1.57$. Since this value exceeds unity, it indicates that the estimated uncertainties systematically underestimate the true mass resolution. In contrast, the semi-leptonic and fully leptonic channels yield standard deviations of $0.93$ and $0.56$ respectively, implying an overestimation of the true uncertainty. Consequently, we calibrate the uncertainty estimates according to Eq.~\ref{pull calibration}, treating each decay channel independently.

\begin{figure}[h]
    \centering
    \includegraphics[width=0.8\textwidth]{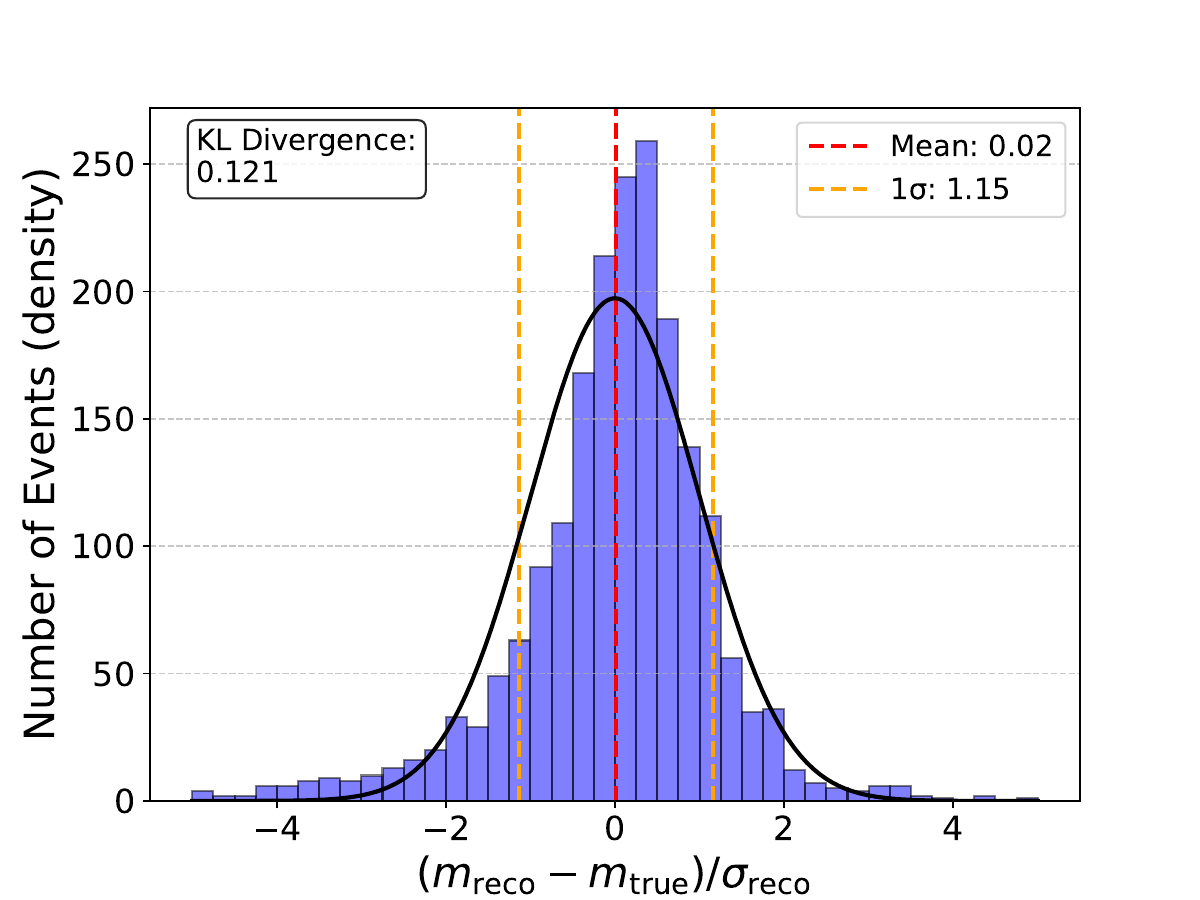}
    \caption{Normalized residual compared to the normal distribution, evaluated on the mixed Higgs-$Z^0$ sample. Normalization was done by subtracting the mean and rescaling by its standard deviation. Their Kullback-Leibler divergence is calculated to be $0.121$.}
    \label{fig:errors}
\end{figure}

We examine the calibrated results by comparing the $z$ distribution with a Gaussian function in Fig.~\ref{fig:errors} for all events and Table~\ref{table:pull_in_different_DMs} for different decay channels. For this comparison, we compute its Kullback--Leibler divergence~\cite{KL_divergence} with respect to a normal distribution $\mathcal{N}(0,1)$, obtaining $D_{KL} = 0.121$. The calibrated $z$ distribution differs from a normal distribution not only in its mean but also in its shape. In particular, it appears visibly narrower than $\mathcal{N}(0,1)$, indicating that the estimated uncertainty $\sigma_\text{reco}$ is, on average, larger than the actual spread of $m_\text{reco} - m_\text{true}$ for events close to the true mass (although this may differ between decay channels). This leads to an overestimation of the uncertainty in the vicinity of the true value. At the same time, since the divergence is relatively small, the uncertainty estimate still provides a qualitatively reasonable description of the mass resolution.

\begin{table}[h]
\centering
\caption{Summary of the normalized residual (pull) for different decay channels.
Mean and standard deviation are reported for each decay channel. Had-had denotes the fully hadronic decay channel, had-lep the semi-leptonic channel, and lep-lep the fully leptonic channel. KL divergence refers to the Kullback–Leibler divergence between the pull distribution and the standard normal distribution $\mathcal{N}(0, 1)$.}
\label{table:pull_in_different_DMs}
\begin{tabular}{|c|ccc|}
\hline
Process & Pull mean & Pull std. dev. & KL divergence \\
\hline
Higgs (had-had) & $-0.20$ & $1.29$ & $0.197$ \\
Higgs (had-lep) & $-0.17$ & $1.06$ & $0.141$ \\
Higgs (lep-lep) & $-0.17$ & $0.94$ & $0.077$ \\
\hline
$Z^0$ (had-had) & $0.30$ & $0.99$ & $0.122$ \\
$Z^0$ (had-lep) & $-0.01$ & $0.90$ & $0.098$ \\
$Z^0$ (lep-lep) & $0.03$ & $0.78$ & $0.139$ \\
\hline
\end{tabular}
\end{table}

\begin{figure}[h]
    \centering
    \includegraphics[width=0.8\textwidth]{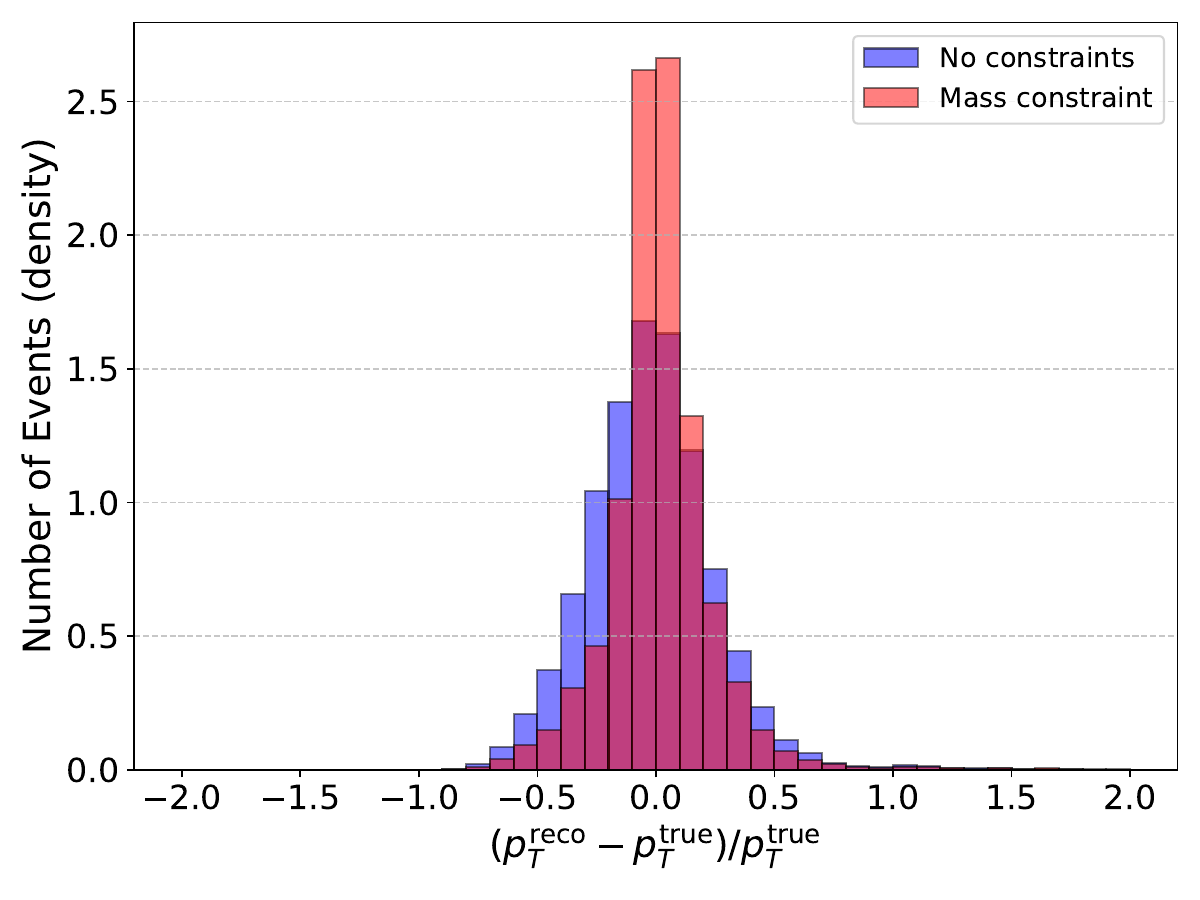}
    \caption{Relative resolution of the reconstructed $p_T$ of taus without mass constraint (blue) and with constraint given by normal distribution with $\sigma = 7.2$ GeV (red). The algorithm was evaluated on the sample coming from the Higgs decay.}
    \label{fig:pt with constraints}
\end{figure}

We have also evaluated the impact of the additional mass constraint on the reconstruction of the $p_T$ of individual $\tau$ leptons.
Fig. \ref{fig:pt with constraints} shows the pulls for the $\tau$ $p_T$ reconstructed with and without the mass constraint. Mass constraint significantly improves the efficiency of $p_T$ reconstruction of individual tau.

After applying the mass constraint, the reconstructed $Z^0$ mass peaks at $m_{Z^0}^\text{constr} = 117$ GeV with a width of $\sigma_{Z^0} = 17$ GeV, illustrating the intrinsic bias introduced when the mass observable is replaced by a kinematic constraint to improve the resolution of the $p_T$ reconstruction.


\subsection{Discussion}

To quantify the results, we compare them with \cite{SVfit} using its fastest approach (cSVfit).

In terms of resolution, we obtained a relative reconstruction resolution of $0.23-0.27$ for the $Z^0$ and $0.24-0.26$ for the Higgs boson (the ranges reflect results separated into different decay channels). For comparison, cSVfit reports relative resolutions of $0.081$--$0.264$ and $0.079$--$0.268$, respectively. Given that the present work is based on simplified simulations, we consider these results to be comparable, with only a small loss in resolution in the case of our algorithm.

We observe that our algorithm achieves a comparable performance across different decay channels, whereas cSVfit exhibits improved performance in hadronic channels. This behaviour is expected, since in hadronic decays the $\overline{E}_T^{\text{miss}}$ is smaller (as only one neutrino escapes the system), and its impact on the reconstruction is therefore reduced. We attribute the comparable performance observed in our results to the relatively weak $\overline{E}_T^{\text{miss}}$ resolution in the simplified simulations used in this study.

For heavy scalar resonances, we observe that our reconstruction remains fairly accurate for masses up to $300$ GeV. For heavier resonances, the reconstructed masses worsen, with a consistent underestimation. This behaviour is likely caused by a combination of poorer $\overline{E}_T^\text{miss}$ resolution in the high-mass region, limitations of our dataset based on simplified Delphes simulations, and stronger regularization applied to stabilize the results.

When comparing these results with cSVfit, we observe similar behaviour in their algorithm. They report a constant underestimation of the reconstructed $A$ mass, with a mean relative difference of $7.2\%$ in the regularized version of the algorithm (and a mean $2.6\%$ mass overestimation in the unregularized version). We conclude that such underestimation is partially expected and may vary depending on the dataset. One should account for these limitations when performing searches for heavy resonances.

Regarding time efficiency, our algorithm can process one event in approximately $5$ ms in a single-thread evaluation. Compared with cSVfit, which reports processing times of $0.34$--$0.77$ s per event, this corresponds to roughly a factor of $100$ speed-up in the case of our algorithm. After comparing the machines used—AMD Opteron 6320 processors (16 logical cores, x86\_64 architecture, 2.8 GHz) in our case and a 2.30 GHz Intel\textsuperscript{\textregistered} Xeon\textsuperscript{\textregistered} E5-2695 v3 for cSVfit -- we expect only a small performance difference due to hardware, slightly favouring the machine used in \cite{SVfit}. This represents the main advantage of our approach, which in practical applications may offset the slightly worse resolution performance.

\section{Summary}

A fast version of a matrix-element-based algorithm for reconstructing the di-tau invariant mass has been developed. While its mass resolution is slightly worse than that of the cSVfit algorithm, it is significantly faster. Moreover, access to the full likelihood map in the space of energy fractions carried by the visible products of the $\tau$ decay allows for the straightforward inclusion of additional constraints, such as mass constraints, and enables an event-by-event estimation of the reconstructed mass uncertainty.

\clearpage

\section*{CRediT authorship contribution statement}
\textbf{Artur Kalinowski}: Conceptualization, Methodology, Software, Validation, Writing – review \& editing.

\textbf{Wiktor Matyszkiewicz}: Methodology, Software, Validation, Formal analysis, Data curation, Writing – original draft.

\section*{Declaration of competing interest(s)}

The authors declare that they have no known competing financial interests or personal relationships that could have appeared to influence the work reported in this paper.

\section*{Acknowledgements}

We would like to express our sincere gratitude to our colleagues from the University of Warsaw and the National Centre for Nuclear Research (NCBJ) for their valuable support throughout this work. In particular, we are indebted to Michał Bluj for his insightful advice and to Krzysztof Doroba for his careful proofreading. We also extend our thanks to the entire CMS Collaboration.

\section*{Data and Code Availability}

The fastMTT source code is available at GitHub (\url{https://github.com/WiktorMat/FastMTT/tree/v-1.0}) and permanently archived on Zenodo~\cite{Github}. 
The Monte Carlo datasets used for the performance studies (Higgs and $Z^0$ boson samples) are available in the same repository.

\section*{Declaration of generative AI and AI-assisted technologies 
in the writing process}

During the preparation of this work, the authors used ChatGPT from OpenAI, in order to improve spelling, grammar, and clarity of the text. After using this service, authors reviewed and edited the content as needed and take full responsibility for the content of the published article.

\clearpage

\appendix

\section{Kinematical limits for tau decay} \label{appendix: kinematical limits}

We derive the kinematic limits in the laboratory (LAB) frame. The tau lepton decay products are divided into invisible products (neutrinos) and visible products (hadrons or charged leptons): 
$\tau \rightarrow \text{vis} + \text{inv}$. Tau has energy $E_\tau$, total three-momentum $p_\tau$, and mass $m_\tau$. Visible products have energy $E_\text{vis}$, total three-momentum $p_\text{vis}$ and mass $m_\text{vis}$, and neutrinos have $E_\nu, p_\nu$ and $m_{\nu\nu}$ respectively. The $x$ is defined as a fraction of energy carried by visible products:

\begin{equation}
    x = \frac{E_\text{vis}}{E_\tau} \label{x definition}
\end{equation}

and respectively:
\begin{equation}
    1-x = \frac{E_\nu}{E_\tau} \label{1-x definition}
\end{equation}

The limits on the $x$ values can be established from the configuration when the visible system moves either in the same or in the opposite direction as the tau lepton in the LAB and tau rest frames. Four-vectors components in the rest frame are denoted with the asterisk, while the laboratory frame components are denoted with the LAB superscript.

\subsection{Hadronic tau decay.}

We shall start with the hadronic decay of tau, where only one, massless, 
neutrino is present. Then, from the conservation of energy:

\begin{equation}E^*_\text{vis} = m_\tau - E_\nu^* = m_\tau - p_\text{vis}^*
\end{equation}

After squaring both sides, we obtain:

\begin{equation}
    p^{2*}_\text{vis} + m^2_\text{vis} = 
    m^2_\tau + p^{2*}_\text{vis} - 2m_\tau p_\text{vis}^* \label{energy_vis_comparison}
\end{equation}

The  expressions for $p_\text{vis}^*$ and $E_\text{vis}^*$ presents as follows:

\begin{align}
 p_\text{vis}^* = \frac{m_\tau^2 - m^2_\text{vis}}{2m_\tau},\\
 E_\text{vis}^* = \frac{m_\tau^2 + m^2_\text{vis}}{2m_\tau}.
\end{align}

The visible products' four-momentum has to be boosted to the LAB frame. We will use the ultrarelativistic $\tau$ lepton limit: $\beta \approx 1$. 

The lower bound on $x$ is obtained for the case when $p_\text{vis}$ is opposite to the boost direction, i.e., when the visible products are emitted in the opposite direction to the tau lepton:

\begin{equation}
    E^\text{LAB}_\text{vis, min} = \gamma (E_\text{vis}^* - p_\text{vis}^*) = 
    \gamma\frac{m^2_\text{vis}}{m_\tau},
\end{equation}

which, after applying $\gamma = \frac{E_\tau}{m_\tau}$, gives:

\begin{equation}
    x_{\text{min}} = \frac{E^\text{LAB}_\text{vis, min}}{E^\text{LAB}_\tau} = 
    \frac{m^2_\text{vis}}{m^2_\tau}.
\end{equation}

Similarly, we obtain an upper limit by considering the same sign of 
momentum and the boost:

\begin{equation}
    E^\text{LAB}_\text{vis, max} = 
    \gamma (E_\text{vis}^* + p_\text{vis}^*) = E_\tau.
\end{equation}

Which translates to:
\begin{equation}
    x_{\text{max}} = \frac{E^\text{LAB}_\text{vis, max}}{E^\text{LAB}_\tau} =
    1
\end{equation}

The limits for the $x$ in the case of hadronic decay are:
\begin{equation}
    \frac{m^2_\text{vis}}{m^2_\tau} <x < 1
\end{equation}

\subsection{Leptonic tau decay.}

\subsubsection{Fraction of energy carried by visible products}

In the case of leptonic decay, the visible products are muons or electrons, with significantly smaller mass than the tau lepton mass; therefore, we can approximate the visible products' mass as zero, $m_\text{vis} = 0$. The energy of the visible products in the rest frame is given by:
 \begin{equation}
        \frac{m^2_\text{vis}}{m^2_\tau}\simeq 0 <x < 1
\end{equation}   

\subsubsection{Invariant mass of neutrino system}

In the case of leptonic decay, we have two neutrinos, which constitute a massive system with invariant mass $m_{\nu\nu}$. Using the conservation of momentum in the tau rest frame, $p^{*}_\nu = p_\text{vis}^*$, we can write:

\begin{equation}
    E^{2*}_\nu = m^2_{\nu \nu} + p^{2*}_\nu = m^2_{\nu \nu} + p^{2*}_\text{vis}.
\end{equation}

We apply this equation to the energy-momentum relation for $E_\text{vis}^*$ and energy conservation, following reasoning introduced in equation \ref{energy_vis_comparison}:

\begin{equation}
    m^2_\tau + m^2_{\nu\nu}+p^{2*}_\text{vis} - 2m_\tau \sqrt{p^{2*}_\text{vis}+m^2_{\nu\nu}} = 
    p^{2*}_\text{vis} + m^{2*}_\text{vis}.
\end{equation}

The $m_\text{vis}$ in the electron or muon mass, therefore term $m_{\tau}^{2} -  m_\text{vis}^2$ can be approximated as $m_\tau^2$.
We get then:

\begin{equation}
    \begin{split}
        m^2_\tau + m^2_{\nu\nu} = 2m_\tau \sqrt{p^{2*}_\text{vis}+m^2_{\nu\nu}} & \\
        m^4_\tau + m^4_{\nu\nu} + 2m^2_\tau m^2_{\nu\nu} = 4m^2_\tau (p^{2*}_\text{vis}+m^2_{\nu\nu}) & \\
        m^4_\tau + m^4_{\nu\nu} - 2m^2_\tau m^2_{\nu\nu} = 4m^2_\tau p^{2*}_\text{vis} & \\
        m^2_\tau - 2m^2_{\nu\nu} = 2m^2_\tau p^{2*}_\text{vis} & \\
        m^2_{\nu\nu} = m^2_\tau - 2m_\tau p_\text{vis}^* & \\
        m^2_{\nu\nu} = m^2_\tau(1 - 2p_\text{vis}^*/m_\tau)
    \end{split}
\end{equation}

We need to express the $p_\text{vis}^*$ in terms of the $x$.
We boost the result into the laboratory frame, again using the ultrarelativistic limit $\beta \approx 1$. The smallest value of the $p_\text{vis}^*$ (leading to the largest value of the $m_{\nu\nu}$) is obtained for the case when the visible products are emitted in the direction to the boost. Therefore, we have:

\begin{equation}
    \begin{split}
    E_\text{vis}^\text{LAB} \simeq \gamma (E_\text{vis}^* + p_\text{vis}^*) \simeq
    \gamma  (p_\text{vis}^* + p_\text{vis}^*) = 2 \gamma p_\text{vis}^* & \\
    2p_\text{vis}^* = \frac{E_\text{vis}m_\tau}{E_\tau} = x m_\tau & \\
    \frac{2p_\text{vis}^*}{m_\tau} = x
    \end{split}
\end{equation}

Finally we get the upper limit for the $m_{\nu\nu}^{2}$:

\begin{equation}
    m^2_{\nu \nu} = m_\tau^2(1 - x).
\end{equation}

The lower bound for $m_{\nu\nu}$ is equal to zero, as this is the configuration when all the neutrinos are emitted in the same direction. Finally, we get the limits for the invariant mass of the neutrino system:
\begin{equation}
    0 < m_{\nu \nu} < m_\tau \sqrt{1 - x}.
\end{equation}

\section{Likelihood separation} \label{appendix: likelihood separation}

To assess the contribution of each component of our likelihood ansatz, we construct two simplified variants of the algorithm.

First, we consider a version that uses only the transfer function for $\overline{E}_T^{\rm miss}$, effectively assuming that the phase–space factor is constant:

\begin{equation}
    \mathcal{L}(\text{data} |x_1, x_2) = \mathcal{N} \,
    W(\text{data}|x_1, x_2).
\end{equation}

Second, we evaluate the complementary version in which the likelihood contains only the phase–space term:

\begin{equation}
    \mathcal{L}(\text{data} | x_1, x_2) = \mathcal{N} \,
    \int d\Phi_n.
\end{equation}

\begin{figure}[h]
    \centering
    \includegraphics[width=\textwidth]{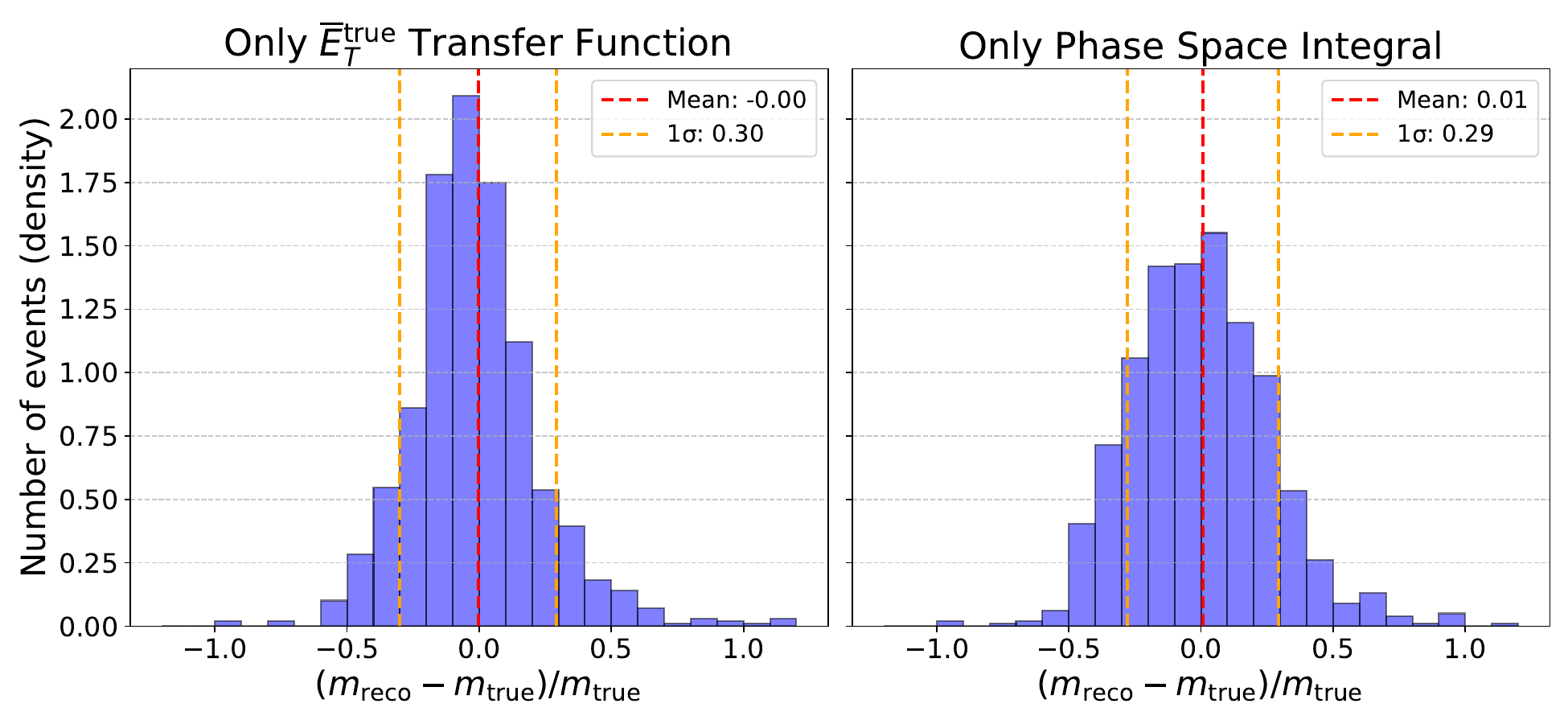}
    \caption{Invariant mass reconstruction of the di-$\tau$ system using only the $\overline{E}_T^{\rm miss}$ transfer function (left) and only the phase-space integration (right). The reconstruction performance is evaluated using a Higgs boson sample.}
    \label{fig:appendix_comparison}
\end{figure}

The resulting mass resolutions $(m_\text{reco} - m_\text{true}) / m_\text{true}$ are summarised in \ref{fig:appendix_comparison}.

The transfer–function–only variant yields a mean of $-0.003$ and a standard deviation of $0.30$. The phase–space–only estimator performs similarly, with a mean of $0.01$ and a slightly smaller spread of $0.29$. For comparison, the full algorithm, which combines both ingredients, achieves a comparable mean of $-0.03$ but a noticeably reduced width of $0.24$ (Figure \ref{fig:mass resolution}).

These results confirm that each component independently carries non-trivial information about the true mass, producing a correctly centered but relatively broad likelihood profile. When combined, their information adds constructively, leading to an approximately 15–20\% improvement in resolution. This behaviour is consistent with the expectation for two moderately correlated likelihood contributions that both peak near the true mass, and supports the empirical validity of our factorized ansatz.

\end{document}